\documentclass[11pt]{article}
\usepackage[left=1in,top=1in,right=1in,bottom=1in]{geometry}
\usepackage{times}

\usepackage{graphicx}
\usepackage{amsmath}
\usepackage{amsthm}
\usepackage{booktabs}
\usepackage{algorithm}
\usepackage{algorithmic}
\usepackage{silence}
\usepackage{amssymb}
\usepackage{verbatim}
\usepackage{url}
\allowdisplaybreaks

\usepackage{siunitx}
\allowdisplaybreaks

\usepackage{placeins}

\newtheorem{definition}{Definition}

\begin{document}

\title{Computing Evolutionarily Stable Strategies in Multiplayer Games}

\author{
Sam Ganzfried\\
Ganzfried Research, Cornell University\\
\texttt{sam.ganzfried@gmail.com}
}

\date{\vspace{-5ex}}

\maketitle

\begin{abstract}
We present an algorithm for computing evolutionarily stable strategies in symmetric normal-form games with three or more players.
\end{abstract}

\section{Introduction}
\label{se:intro}
While Nash equilibrium has emerged as the standard solution concept in game theory, it is often criticized as being too weak: often games contain multiple Nash equilibria (sometimes even infinitely many), and we want to select one that satisfies other natural properties. For example, one popular concept that refines Nash equilibrium is evolutionarily stable strategy (ESS). A mixed strategy in a two-player symmetric game is an evolutionarily stable strategy if it is robust to being overtaken by a mutation strategy. Formally, mixed strategy $\mathbf{x}^*$ is an ESS if for every mixed strategy $\mathbf{x}$ that differs from $\mathbf{x}^*$, there exists $\epsilon_0 = \epsilon_0(\mathbf{x}) > 0$ such that, for all $\epsilon \in (0,\epsilon_0)$, 
\begin{equation*}
(1-\epsilon)u_1(\mathbf{x},\mathbf{x}^*) + \epsilon u_1(\mathbf{x},\mathbf{x}) <  (1-\epsilon)u_1(\mathbf{x}^*,\mathbf{x}^*)+\epsilon u_1(\mathbf{x}^*,\mathbf{x}).
\end{equation*}
From a biological perspective, we can interpret $\mathbf{x}^*$ as a distribution among ``normal'' individuals within a population, and consider a mutation that makes use of strategy $\mathbf{x}$, assuming that the proportion of the mutation in the population is $\epsilon$. In an ESS, the expected payoff of the mutation is smaller than the expected payoff of a normal individual, and hence the proportion of mutations will decrease and eventually disappear over time, with the composition of the population returning to being mostly $\mathbf{x}^*$. An ESS is therefore a resident mixed strategy that is immune to being overtaken by mutations. ESS was initially proposed by mathematical biologists motivated by applications such as population dynamics (e.g., maintaining robustness to mutations within a population of humans or animals)~\cite{Maynard73:Logic,Maynard82:Evolution}. A common example game is the 2x2 game where strategies correspond to an ``aggressive'' Hawk or a ``peaceful'' Dove strategy. A paper has proposed a similar game in which an aggressive malignant cell competes with a passive normal cell for biological energy, which has applications to cancer eradication~\cite{Dingli09:Cancer}. 

While Nash equilibrium is defined for general multiplayer games, ESS is traditionally studied in the context of symmetric two-player games. ESS is a refinement of Nash equilibrium. In particular, if $\mathbf{x}^*$ is an ESS, then $(\mathbf{x}^*,\mathbf{x}^*)$ (i.e., the strategy profile where both players play $\mathbf{x}^*$) is a (symmetric) Nash equilibrium~\cite{Maschler13:Game}. Of course the converse is not necessarily true (not every symmetric Nash equilibrium is an ESS), or else ESS would be a trivial refinement. In fact, ESS is not guaranteed to exist in finite games (while Nash equilibrium is guaranteed to exist in all finite games). For example, while rock-paper-scissors has a mixed strategy Nash equilibrium (which puts equal weight on all three actions), it has no ESS~\cite{Maschler13:Game} (that work considers a version where payoffs are 1 for a victory, 0 for loss, and $\frac{2}{3}$ for a tie).

There exists a polynomial-time algorithm for computing Nash equilibrium (NE) in two-player zero-sum games, while for two-player non-zero-sum and multiplayer games computing an NE is PPAD-hard and it is widely conjectured that no efficient (polynomial-time) algorithm exists. However, several algorithms have been devised that perform well in practice. The problem of deciding whether a game has an ESS was shown to be both NP-hard and co-NP-hard and also to be contained in $\Sigma ^P _2$ (the class of decision problems that can be solved in nondeterministic polynomial time given access to an NP oracle)~\cite{Etessami08:Computational}. Subsequently it was shown that the exact complexity is $\Sigma ^P _2$-complete~\cite{Conitzer13:Exact}, and even more challenging for more than two players~\cite{Blanc21:Computational}. Note that this result is for determining whether an ESS exists (as discussed above there exist games which have no ESS), not for the complexity of computing an ESS in games for which one exists. Thus, computing an ESS is significantly more difficult than computing an NE, which is not surprising since it is a refinement. Several approaches have been proposed for computing ESSs in two-player games~\cite{Haigh75:Game,Abakuks80:Conditions,Broom13:Game-Theoretical,Bomze92:Detecting,Mcnamara97:General}. However, we are not aware of approaches for computing ESSs in games with 3+ players.

Many evolutionary models of tumor ecology involve frequency-dependent interactions among three or more cancer cell phenotypes (e.g., \cite{Basanta08:Studying,Basanta12:Investigating,Archetti13:Evolutionary,Laruelle23:Effects}). These systems are naturally formulated as symmetric $n$-player games, and ESS provides a biologically meaningful stability concept. Motivated by this, we develop an algorithm for computing evolutionarily stable strategies in multiplayer symmetric normal-form games. Evolutionarily stable strategies in multiplayer symmetric games have been studied in the evolutionary game theory literature, beginning with early work by Palm \cite{Palm84:Evolutionary} and subsequently developed by Broom, Cannings, and Vickers \cite{Broom97:Multi}, among others.

A \emph{normal-form game} consists of a finite set of players $N = \{1,\ldots,n\}$, a finite set of pure strategies $S_i$ for each player $i$, and a real-valued utility for each player for each strategy vector (aka \emph{strategy profile}), $u_i : \times_j S_j \rightarrow \mathbb{R}$. In a \emph{symmetric normal-form game}, all strategy spaces $S_i$ are equal and the utility functions satisfy the following symmetry condition: for every player $i \in N$, pure strategy profile $(s_1,\ldots,s_n) \in S^n$, and permutation $\pi$ of the players,
$$u_i(s_1,\ldots,s_n) = u_{\pi(i)}(s_{\pi^{-1}(1)},\ldots,s_{\pi^{-1}(n)}).$$
This allows us to remove the player index of the utility function and just write 
$u(s_1,\ldots,s_n),$ where it is implied that the utility is for player 1 (we can simply permute the players to obtain the utilities of the other players). We will still write $u_i$ for notational convenience, but note that only a single utility function must be specified which will apply to all players.

Let $\Sigma_i$ denote the set of mixed strategies of player $i$ (probability distributions over elements of $S_i$). If players follow mixed strategy profile
$\mathbf{x} = (\mathbf{x}^{(1)},\ldots,\mathbf{x}^{(n)}),$ where $\mathbf{x}^{(i)} \in \Sigma_i$, the expected payoff to player $i$ is
$$u_i(\mathbf{x}^{(1)},\ldots,\mathbf{x}^{(n)}) = \sum_{s_1,\ldots,s_n \in S} x^{(1)}_{s_1} \cdots x^{(n)}_{s_n}u_i(s_1,\ldots,s_n).$$ 
We write $u_i(\mathbf{x}) = u_i(\mathbf{x}^{(i)},\mathbf{x}^{(-i)})$,
where $\mathbf{x}^{(-i)}$ denotes the vector of strategies of all players except $i$.
If all players follow the same mixed strategy $\mathbf{x}$, then for all players $i$ we have
$$u_i(\mathbf{x}) = u_i(\mathbf{x},\ldots,\mathbf{x}) = \sum_{s_1,\ldots,s_n \in S} x_{s_1} \cdots x_{s_n} u_i(s_1,\ldots,s_n).$$ 

\begin{definition}
\label{de:ne}
A mixed strategy profile $\mathbf{x}^*$ is a Nash equilibrium if for each player $i \in N$ and for each mixed strategy $\mathbf{x}^{(i)} \in \Sigma_i$:
$u_i(\mathbf{x}^{*(i)},\mathbf{x}^{*(-i)}) \geq u_i(\mathbf{x}^{(i)},\mathbf{x}^{*(-i)}).$
\end{definition}

\begin{definition}
\label{de:ne-sym}
A mixed strategy profile $\mathbf{x}^*$ in a symmetric normal-form game is a symmetric Nash equilibrium if it is a Nash equilibrium and: 
$\mathbf{x}^{*(1)} = \mathbf{x}^{*(2)} = \cdots = \mathbf{x}^{*(n)}.$
\end{definition}

\begin{definition}[Broom, Cannings, and Vickers \cite{Broom97:Multi}]
\label{de:ess}
A mixed strategy $\mathbf{x}^* \in \Sigma_1$ is \emph{evolutionarily stable} in a symmetric normal-form game if for each mixed strategy $\mathbf{x} \neq \mathbf{x}^*$ there exists an integer $r \in \{0,\ldots,n-1\}$ such that
\[
u_1(\mathbf{x}^*,\underbrace{\mathbf{x},\ldots,\mathbf{x}}_{r},
\underbrace{\mathbf{x}^*,\ldots,\mathbf{x}^*}_{n-1-r})
>
u_1(\underbrace{\mathbf{x},\ldots,\mathbf{x}}_{r+1},
\underbrace{\mathbf{x}^*,\ldots,\mathbf{x}^*}_{n-1-r}),
\]
and for every $j<r$,
\[
u_1(\mathbf{x}^*,\underbrace{\mathbf{x},\ldots,\mathbf{x}}_{j},
\underbrace{\mathbf{x}^*,\ldots,\mathbf{x}^*}_{n-1-j})
=
u_1(\underbrace{\mathbf{x},\ldots,\mathbf{x}}_{j+1},
\underbrace{\mathbf{x}^*,\ldots,\mathbf{x}^*}_{n-1-j}).
\]
\end{definition}

For the special case of $n = 3$ players, a mixed strategy $\mathbf{x}^*$ is evolutionarily stable if for every mixed strategy $\mathbf{x} \neq \mathbf{x}^*$ exactly one of the following conditions holds:

\begin{enumerate}
\item $u_1(\mathbf{x}^*,\mathbf{x}^*,\mathbf{x}^*) > u_1(\mathbf{x},\mathbf{x}^*,\mathbf{x}^*)$,

\item $u_1(\mathbf{x}^*,\mathbf{x}^*,\mathbf{x}^*) = u_1(\mathbf{x},\mathbf{x}^*,\mathbf{x}^*)$ and
$u_1(\mathbf{x}^*,\mathbf{x},\mathbf{x}^*) > u_1(\mathbf{x},\mathbf{x},\mathbf{x}^*)$,

\item $u_1(\mathbf{x}^*,\mathbf{x}^*,\mathbf{x}^*) = u_1(\mathbf{x},\mathbf{x}^*,\mathbf{x}^*)$,
$u_1(\mathbf{x}^*,\mathbf{x},\mathbf{x}^*) = u_1(\mathbf{x},\mathbf{x},\mathbf{x}^*)$, and
$u_1(\mathbf{x}^*,\mathbf{x},\mathbf{x}) > u_1(\mathbf{x},\mathbf{x},\mathbf{x})$.
\end{enumerate}

It has been proven that every symmetric normal-form game has at least one symmetric Nash equilibrium~\cite{Nash51:Non}. It is clear from Definition~\ref{de:ess} that every evolutionarily stable strategy (ESS) in a symmetric normal-form game must be a symmetric Nash equilibrium (SNE). Two-player SNEs with a prescribed support satisfy linear constraints. If that support contains multiple SNEs, none can be evolutionarily stable. Consequently, testing one SNE per support is sufficient when searching for ESSs, even in degenerate games. With three or more players, the corresponding constraints are polynomial, and an ESS can share its support with another SNE that is not evolutionarily stable. Our algorithm therefore continues searching each support after finding an initial SNE.

\section{Algorithm}
\label{se:algorithm}
In this section we present our main algorithm for computing ESSs in symmetric normal-form games. We present our
algorithm just for the three-player case and discuss how it can be extended to $n$ players. The algorithm works by enumerating supports. For each support, it first computes a symmetric Nash equilibrium (SNE) if one exists and tests whether it is an ESS. It then repeatedly searches the same support for an additional SNE that is sufficiently far from all SNEs already found on that support. Thus, the algorithm can find multiple isolated SNEs on a single support. The procedures for computing an SNE and testing whether an SNE is an ESS involve solving nonconvex quadratically-constrained programs.

Algorithm~\ref{al:main} presents pseudocode for the main algorithm. The input is a payoff tensor $\mathbf{A}$, where $A[i,j,k]$ is the payoff to player 1 when player 1 plays pure strategy $i,$ player 2 plays pure strategy $j$, and player 3 plays pure strategy $k.$ The algorithm also uses several numerical parameters whose values are given in Table~\ref{ta:parameters}. For each support $\mathbf{T}$, Algorithm~\ref{al:sne-support} finds an initial SNE if one exists. After each SNE $\mathbf{x}$ is tested by Algorithm~\ref{al:isess}, Algorithm~\ref{al:sne-maxdist} searches for another SNE on $\mathbf{T}$ that is separated from all previously found SNEs on that support. The search continues until no further SNE is found at the specified separation threshold, or until a numerical solve is unresolved. For clarity, Algorithm~\ref{al:main} omits numerical-status bookkeeping: in the implementation, any unresolved SNE search or ESS classification is recorded and the overall run is reported as incomplete.

\begin{algorithm}[!ht]
\caption{Compute ESSs in a 3-player symmetric normal-form game}
\label{al:main}
\begin{algorithmic}[1]
\REQUIRE Payoff tensor $\mathbf{A}[i,j,k]$, number of strategies $K$
\REQUIRE Tolerances $\epsilon_s,\epsilon_p,\delta,\epsilon_{\mathrm{dist}}$
\ENSURE Set $\mathrm{ESS\_set}$
\STATE $\mathrm{ESS\_set} \leftarrow \emptyset$
\FOR{$m=1$ to $K$}
    \FOR{each subset $\mathbf{T} \subseteq \{0,\ldots,K-1\}$ with $|\mathbf{T}|=m$}
        \STATE $\mathcal{L}_{\mathbf{T}} \leftarrow \emptyset$
        \STATE $(status,\mathbf{x}) \leftarrow
        \textsc{SNE\_Support}(\mathbf{A},\mathbf{T},\epsilon_s)$
        \WHILE{$status=\mathrm{FEASIBLE}$}
            \STATE $\mathcal{L}_{\mathbf{T}} \leftarrow
            \mathcal{L}_{\mathbf{T}} \cup \{\mathbf{x}\}$
            \IF{\textsc{IsESS}$(\mathbf{A},\mathbf{x},\epsilon_p,\delta)=\mathrm{true}$}
                \STATE $\mathrm{ESS\_set} \leftarrow
                \mathrm{ESS\_set}\cup\{\mathbf{x}\}$
            \ENDIF
            \STATE $(status,\mathbf{x}) \leftarrow
            \textsc{SNE\_MaxDist}
            (\mathbf{A},\mathbf{T},\mathcal{L}_{\mathbf{T}},
            \epsilon_s,\epsilon_{\mathrm{dist}})$
        \ENDWHILE
    \ENDFOR
\ENDFOR
\RETURN $\mathrm{ESS\_set}$
\end{algorithmic}
\end{algorithm}

\begin{algorithm}[!ht]
\caption{\textsc{SNE\_Support}$(\mathbf{A},\mathbf{T},\epsilon_s)$}
\label{al:sne-support}
\begin{algorithmic}[1]

\STATE Create a QCP with variables $p_\ell \ge \epsilon_s$ for $\ell \in \mathbf{T}$
\STATE Add constraint $\sum_{\ell \in \mathbf{T}} p_\ell = 1$
\STATE Define $i0 = T[0]$

\STATE Define $g_i(\mathbf{p}) = \sum_{j \in \mathbf{T}} \sum_{k \in \mathbf{T}}
A[i,j,k]\, p_j p_k$.

\FOR{$i \in \mathbf{T}$, $i \neq i0$}
    \STATE Add constraint $g_i(\mathbf{p}) = g_{i0}(\mathbf{p})$
\ENDFOR

\FOR{$i \notin \mathbf{T}$}
    \STATE Add constraint $g_i(\mathbf{p}) \le g_{i0}(\mathbf{p})$
\ENDFOR

\STATE Solve the QCP
\IF{proven infeasible} \STATE \textbf{return} (INFEASIBLE, $\bot$) \ENDIF
\IF{the solve or recomputed feasibility check is unresolved} \STATE \textbf{return} (UNRESOLVED, $\bot$) \ENDIF

\STATE Construct $\mathbf{x}$ by $x_i=p_i$ for $i\in\mathbf{T}$ and $x_i=0$ otherwise
\RETURN (FEASIBLE, $\mathbf{x}$)

\end{algorithmic}
\end{algorithm}

\begin{algorithm}[!ht]
\caption{\textsc{SNE\_MaxDist}$(\mathbf{A},\mathbf{T},\mathcal{L}_{\mathbf{T}},\epsilon_s,\epsilon_{\mathrm{dist}})$}
\label{al:sne-maxdist}
\begin{algorithmic}[1]

\REQUIRE Payoff tensor $\mathbf{A}$; support $\mathbf{T} \subseteq \{0,\ldots,K-1\}$;
previously found SNEs $\mathcal{L}_{\mathbf{T}}=\{\mathbf{x}^{(1)},\ldots,\mathbf{x}^{(r)}\}$; support tolerance $\epsilon_s > 0$; squared-distance tolerance $\epsilon_{\mathrm{dist}} > 0$.

\STATE Create the SNE QCP constraints from Algorithm~\ref{al:sne-support} with variables $p'_\ell$ for $\ell\in\mathbf{T}$.
\STATE Create variable $t\in[0,2]$.
\FOR{$q=1,\ldots,r$}
    \STATE Add constraint $\sum_{\ell\in\mathbf{T}}(p'_\ell-x^{(q)}_\ell)^2 \ge t$.
\ENDFOR
\STATE Maximize $t$ and solve the QCQP.
\STATE Let $\bar t$ be the solver's upper bound on the optimal value and let $\mathbf{x}$ be a returned, residual-validated SNE; if none is returned, return $(\textsc{UNRESOLVED},\bot)$.
\IF{$\min_q\|\mathbf{x}-\mathbf{x}^{(q)}\|_2^2 > \epsilon_{\mathrm{dist}}$}
    \RETURN (FEASIBLE,$\mathbf{x}$)
\ELSIF{$\bar t \le \epsilon_{\mathrm{dist}}$}
    \RETURN (EXHAUSTED,$\bot$)
\ELSE
    \RETURN (UNRESOLVED,$\bot$)
\ENDIF

\end{algorithmic}
\end{algorithm}

Algorithm~\ref{al:sne-support} tests whether the game has an SNE with given support $\mathbf{T}$ by creating and solving a quadratically-constrained feasibility program (QCP). The variables $p_\ell$ represent the non-negative strategy probabilities over the elements of $\mathbf{T}$. We define $g_i(\mathbf{p})$ to be the expected payoff of playing pure strategy $i$ when the opposing players play $\mathbf{p}.$ Note that $g_i(\mathbf{p})$ is quadratic since it contains products of variables $p_j p_k.$ For each $i \in \mathbf{T},$ we add the constraint $g_i(\mathbf{p}) = g_{i0}(\mathbf{p})$, and for each $i \notin \mathbf{T}$, we add the constraint $g_i(\mathbf{p}) \le g_{i0}(\mathbf{p})$. This ensures that the player is indifferent between all pure strategies in the support and cannot profitably deviate to a strategy outside the support, which is the definition of Nash equilibrium.

A support can contain more than one SNE, so solving Algorithm~\ref{al:sne-support} only once on each support is not sufficient. After finding an initial SNE, we therefore apply Algorithm~\ref{al:sne-maxdist}. Given the set $\mathcal{L}_{\mathbf{T}}$ of SNEs already found on support $\mathbf{T}$, the algorithm finds an SNE that maximizes its minimum squared Euclidean distance from the members of $\mathcal{L}_{\mathbf{T}}$. All previously found SNEs are retained in this set, regardless of whether they pass the ESS test.

If Algorithm~\ref{al:sne-maxdist} returns EXHAUSTED, then at the specified numerical resolution every SNE satisfying the support constraints is within Euclidean distance $\sqrt{\epsilon_{\mathrm{dist}}}$ of an SNE already found. Thus, if the SNEs on a support are finite and pairwise separated by more than this distance, the repeated search finds each of them individually. If a support contains a continuum of SNEs, or distinct equilibria closer than the threshold, the procedure returns separated representatives rather than enumerating every exact equilibrium. In the implementation, a small returned distance is not sufficient to terminate the search; the solver's upper bound must also be at most $\epsilon_{\mathrm{dist}}$. We additionally impose a safety cap of 100 representatives per support and a 300-second limit per solve; reaching either limit is reported as incomplete rather than exhaustion.

Given SNE $\mathbf{x}$, Algorithm~\ref{al:isess} determines whether it is an ESS. 
The multilinear payoff to player~1 is:
$$ U(\mathbf{a},\mathbf{b},\mathbf{c}) = \sum_{i,j,k} A[i,j,k]\, a_i b_j c_k.$$
We first check whether $\mathbf{x}$ is a strict Nash equilibrium. 
It is well known that in symmetric games any strict SNE is an 
ESS~\cite{Maynard73:Logic,Maynard82:Evolution,Hofbauer98:Evolutionary}.
Moreover, among symmetric Nash equilibria, strictness is equivalent to 
being a pure strategy whose chosen action is the unique best response to 
itself~\cite{Maynard82:Evolution,Hofbauer98:Evolutionary}.  
Consequently, whenever $\mathbf{x}$ is pure and satisfies $|BR(\mathbf{x})|=1$,
we may immediately conclude that $\mathbf{x}$ is an ESS. 
Note that $|BR(\mathbf{x})|=1$ implies that $\mathbf{x}$ is pure, so we can simplify this
procedure to just check if $|BR(\mathbf{x})|=1.$ 
A mutant can tie the resident in the initial comparison of Definition~\ref{de:ess} only if it is supported on best responses to $\mathbf{x}$.
Accordingly, the mixed-mutant searches are restricted to $BR(\mathbf{x})$, which numerically includes near-best responses within tolerance $\epsilon_p$.
We next test whether a pure mutant can invade $\mathbf{x}$ by iterating over all pure strategies
$i \in \mathrm{BR}(\mathbf{x})$ such that $\mathbf{e}_i \neq \mathbf{x}$. For each such $i$, we first compare
$U(\mathbf{x},\mathbf{e}_i,\mathbf{x})$ and $U(\mathbf{e}_i,\mathbf{e}_i,\mathbf{x})$.
If the two quantities are equal up to numerical tolerance, then we apply the second-level comparison
$U(\mathbf{x},\mathbf{e}_i,\mathbf{e}_i)$ versus $U(\mathbf{e}_i,\mathbf{e}_i,\mathbf{e}_i)$,
as required by Definition~\ref{de:ess}.
After applying these efficient tests, we apply Algorithms~\ref{al:ess-level1}
and~\ref{al:ess-level2} to determine whether $\mathbf{x}$ can be invaded
by any mixed mutant.

\begin{algorithm}[!ht]
\caption{\textsc{IsESS}$(\mathbf{A},\mathbf{x},\epsilon_p,\delta)$}
\label{al:isess}
\begin{algorithmic}[1]

\STATE Let $\mathbf{e}_i$ be pure strategy placing all mass on action $i$.
\STATE Compute $g_i = U(\mathbf{e}_i,\mathbf{x},\mathbf{x})$ for all $i$
\STATE Compute $v = U(\mathbf{x},\mathbf{x},\mathbf{x})$
\STATE $\mathrm{BR}(\mathbf{x}) = \{\, i : g_i \ge v - \epsilon_p \,\}$

\STATE \textbf{/* Strict NE shortcut */}
\IF{$\mathbf{x}$ is pure and $|\mathrm{BR}(\mathbf{x})| = 1$}
    \RETURN \TRUE
\ENDIF

\STATE \textbf{/* Pure-mutant screen */}
\FOR{each $i \in \mathrm{BR}(\mathbf{x})$ such that $\mathbf{e}_i \neq \mathbf{x}$}
    \STATE Compute $u_1 \gets U(\mathbf{x}, \mathbf{e}_i, \mathbf{x})$
    \STATE Compute $u_2 \gets U(\mathbf{e}_i, \mathbf{e}_i, \mathbf{x})$

    \IF{$u_1 < u_2 - \epsilon_p$}
        \RETURN \FALSE \quad $\triangleright$ pure mutant $i$ is not repelled at level 1
    \ELSIF{$|u_1-u_2| \le \epsilon_p$}
        \STATE Compute $u_3 \gets U(\mathbf{x}, \mathbf{e}_i, \mathbf{e}_i)$
        \STATE Compute $u_4 \gets U(\mathbf{e}_i, \mathbf{e}_i, \mathbf{e}_i)$
        \IF{$u_3 \le u_4 + \epsilon_p$}
            \RETURN \FALSE \quad $\triangleright$ pure mutant $i$ is not repelled at level 2
        \ENDIF
    \ENDIF
\ENDFOR

\STATE $F^\star \gets \textsc{ESS\_LEVEL1}(\mathbf{A},\mathbf{x},\mathrm{BR}(\mathbf{x}),\delta)$
\IF{the level-1 solve is unresolved}
    \RETURN \textsc{UNRESOLVED}
\ELSIF{$F^\star < -\epsilon_p$}
    \RETURN \FALSE \quad $\triangleright$ mixed mutant is not repelled at level 1
\ELSIF{$F^\star > \epsilon_p$}
    \RETURN \TRUE \quad $\triangleright$ all mixed mutants repelled at level 1
\ELSE
    \STATE $G^\star \gets \textsc{ESS\_LEVEL2}(\mathbf{A},\mathbf{x},\mathrm{BR}(\mathbf{x}),\epsilon_p,\delta)$
		\IF{the level-2 solve is unresolved}
			\RETURN \textsc{UNRESOLVED}
    \ELSIF{$G^\star > \epsilon_p$}
        \RETURN \TRUE \quad $\triangleright$ level-1 ties repelled at level 2
    \ELSE
        \RETURN \FALSE \quad $\triangleright$ mixed mutant is not repelled at level 2
    \ENDIF
\ENDIF
\end{algorithmic}
\end{algorithm}

Algorithm~\ref{al:ess-level1} performs the first-level mixed-mutant test.
The variables $y_i$ represent the non-negative strategy probabilities over
the elements of $\mathrm{BR}(\mathbf{x})$. Let $\mathbf{y}$ denote the extension
of this vector to a $K$-dimensional strategy by setting entries not in
$\mathrm{BR}(\mathbf{x})$ to zero. The algorithm minimizes
\[
F(\mathbf{y}) =
U(\mathbf{x},\mathbf{y},\mathbf{x}) -
U(\mathbf{y},\mathbf{y},\mathbf{x}).
\]
If the minimum value is positive (i.e., numerically exceeds $\varepsilon_p$), the candidate passes the first-level mixed-mutant test of Definition~\ref{de:ess}. 
If it is negative (i.e., below $-\varepsilon_p$), the candidate fails this numerical test.
If the value is approximately zero, then
Algorithm~\ref{al:ess-level2} applies the second-level comparison by
minimizing
\[
G(\mathbf{y}) =
U(\mathbf{x},\mathbf{y},\mathbf{y}) -
U(\mathbf{y},\mathbf{y},\mathbf{y})
\]
over mutants for which $F(\mathbf{y})$ is approximately zero.
Numerically, we say that $\mathbf{y} \neq \mathbf{x}$ when
$\|\mathbf{y}-\mathbf{x}\|_2 \ge \delta$.

\begin{algorithm}[!ht]
\caption{\textsc{ESS\_LEVEL1}$(\mathbf{A},\mathbf{x},\mathrm{BR}(\mathbf{x}),\delta)$}
\label{al:ess-level1}
\begin{algorithmic}[1]

\STATE Create a QCQP with variables $y_i \ge 0$ for $i \in \mathrm{BR}(\mathbf{x})$
\STATE Add constraint $\sum_{i \in \mathrm{BR}(\mathbf{x})} y_i = 1$
\STATE Add constraint $\sum_{i \in \mathrm{BR}(\mathbf{x})} (y_i - x_i)^2 \ge \delta^2$
\STATE Define $F(\mathbf{y}) = U(\mathbf{x},\mathbf{y},\mathbf{x}) - U(\mathbf{y},\mathbf{y},\mathbf{x})$
\STATE Solve QCQP minimizing $F(\mathbf{y})$ subject to constraints
\STATE $F^\star \gets$ optimal value of $F(\mathbf{y})$
\RETURN $F^\star$

\end{algorithmic}
\end{algorithm}

\begin{algorithm}[!ht]
\caption{\textsc{ESS\_LEVEL2}$(\mathbf{A},\mathbf{x},\mathrm{BR}(\mathbf{x}),\epsilon_p,\delta)$}
\label{al:ess-level2}
\begin{algorithmic}[1]

\STATE Create a QCQP with variables $y_i \ge 0$ for $i \in \mathrm{BR}(\mathbf{x})$
\STATE Add constraint $\sum_{i \in \mathrm{BR}(\mathbf{x})} y_i = 1$
\STATE Add constraint $\sum_{i \in \mathrm{BR}(\mathbf{x})} (y_i - x_i)^2 \ge \delta^2$
\STATE Define $F(\mathbf{y}) = U(\mathbf{x},\mathbf{y},\mathbf{x}) - U(\mathbf{y},\mathbf{y},\mathbf{x})$
\STATE Add constraint $-\epsilon_p \le F(\mathbf{y}) \le \epsilon_p$
\STATE Define $G(\mathbf{y}) = U(\mathbf{x},\mathbf{y},\mathbf{y}) - U(\mathbf{y},\mathbf{y},\mathbf{y})$
\STATE Solve QCQP minimizing $G(\mathbf{y})$ subject to constraints
\STATE $G^\star \gets$ optimal value of $G(\mathbf{y})$
\RETURN $G^\star$

\end{algorithmic}
\end{algorithm}

The QCP in Algorithm~\ref{al:sne-support} and the QCQPs in
Algorithms~\ref{al:sne-maxdist} and~\ref{al:ess-level1} have nonconvex quadratic constraints, and
Algorithm~\ref{al:ess-level1} also has a nonconvex quadratic objective.
Algorithm~\ref{al:ess-level2} introduces the second-level mixed-mutant
test. For the 3-player case, the objective
$
G(\mathbf{y}) =
U(\mathbf{x},\mathbf{y},\mathbf{y}) -
U(\mathbf{y},\mathbf{y},\mathbf{y})
$
contains cubic terms in $\mathbf{y}$. We reformulate this problem as a QCQP
by introducing auxiliary variables for products of the components of
$\mathbf{y}$, and solve all of these programs using Gurobi's nonconvex
quadratic solver~\cite{Gurobi26:Gurobi}.

The numerical parameters are given in Table~\ref{ta:parameters}. The support tolerance
$\epsilon_s$ requires strategies in an enumerated support to be played with probability
at least $\epsilon_s$; $\epsilon_p$ is the payoff-comparison tolerance; $\delta$ excludes
mutants within Euclidean distance $\delta$ of the resident; and
$\epsilon_{\mathrm{dist}}$ is a squared Euclidean-distance threshold used to distinguish
SNEs found on the same support. In the implementation we set Gurobi's feasibility and
optimality tolerances to $10^{-9}$, the relative optimality gap to zero, and the absolute
optimality gap to $10^{-10}$, and recompute solution residuals with tolerance $10^{-8}$.
For the SNE-search subproblems we additionally set \texttt{NumericFocus=3}. For mixed-mutant programs, a positive returned objective value is not by itself sufficient to establish positivity of the global minimum; we use the solver's lower bound on the optimal objective value, and mark a classification unresolved if the bound or the recomputed feasibility checks do not resolve the comparison.
Proven infeasibility of a mutant-search problem means that
no mutant satisfies that stage's constraints and is not treated as an ESS failure.


Because $\epsilon_s$, $\epsilon_p$, $\delta$, and $\epsilon_{\mathrm{dist}}$ are positive numerical thresholds, the implementation is a finite-tolerance procedure rather than an exact-arithmetic ESS oracle. In particular, SNEs with support probabilities below $\epsilon_s$ may be omitted,
distinct equilibria within squared Euclidean distance $\epsilon_{\mathrm{dist}}$
may be represented by a single point, and the mixed-mutant tests do not search
within Euclidean distance $\delta$ of the resident. In the random-game experiments, ``ESS'' denotes classification by this numerical procedure.

\begin{table}[!ht]
\centering
\caption{Parameter values used in the algorithm.}
\label{ta:parameters}
\begin{tabular}{|*{2}{c|}} \hline
Parameter &Value\\ \hline
$\epsilon_s$ & $10^{-4}$\\  \hline 
$\epsilon_p$ & $10^{-5}$ \\ \hline
$\delta$    & $10^{-2}$ \\ \hline
$\epsilon_{\mathrm{dist}}$ (squared distance) & $10^{-6}$ \\  
\hline
\end{tabular}
\end{table}

While we have presented the algorithm just for the 3-player case, it extends naturally to $n > 3$ players. The main difference is that $g_i(\mathbf{p})$ in
Algorithm~\ref{al:sne-support} will involve a product of $n-1$ variables instead of 2. We can convert these expressions to equivalent
expressions that are quadratic by introducing auxiliary variables. For example for $n = 4$, we can set $p_{12} = p_1 p_2$, $p_{123} = p_{12}p_3$, etc. 
Thus the programs will have more variables, but they will remain quadratic and can be solved as a QCP with Gurobi's nonconvex quadratic solver.

The full ESS test requires additional higher-level comparisons analogous to Algorithm~\ref{al:ess-level2}. For $r=1,\ldots,n-1$, define
\[
H_r(\mathbf{y})=
U(\mathbf{x},\underbrace{\mathbf{y},\ldots,\mathbf{y}}_{r},\underbrace{\mathbf{x},\ldots,\mathbf{x}}_{n-1-r})-
U(\underbrace{\mathbf{y},\ldots,\mathbf{y}}_{r+1},\underbrace{\mathbf{x},\ldots,\mathbf{x}}_{n-1-r}).
\]
The $H_1$ mixed-mutant test minimizes $H_1(\mathbf{y})$, which is quadratic for any number of players.
If the minimum is numerically tied at zero, the next level minimizes $H_2(\mathbf{y})$ subject to all previous comparisons being tied, and so on through $H_{n-1}$. These tests involve higher-degree polynomial objectives and constraints, which can be reformulated as QCQPs using auxiliary product variables. Also note that the number of supports enumerated over in Algorithm~\ref{al:main} remains $2^K - 1$ independently of the number of players $n.$

\section{Experiments}
\label{se:experiments}
Our first set of experiments is on a set of example games given in Appendix~\ref{ap:examples}. The full sets of SNEs and ESSs for these games are summarized in Table~\ref{ta:example-games}. Game 1 contains two distinct SNEs on the full support, $(1/3,2/3)$ and $(1/2,1/2)$, of which only the latter is an ESS. This illustrates why it is necessary to continue searching a support after the first SNE is found. Games 2 and 5 contain a continuum of SNEs on one support; for such supports a finite positive-distance search returns separated representatives rather than every exact equilibrium, and reaching the representative cap is reported as incomplete. The ESS set is empty for Game 2, while Game 5 has the unique ESS $(1,0,0)$. For the remaining example games the SNE sets are finite, and the implementation finds the reference SNE and ESS sets. Game 6 represents frequency-dependent competition among three tumor phenotypes---Proliferators, Producers, and Invasives---where each cell's payoff equals the sum of its pairwise interactions with two neighbors. Proliferators and Invasives each receive the highest payoff when interacting with their own type, while Producers confer no competitive advantage.

\begin{table}[ht]
\centering
\caption{Summary of SNEs and ESSs for the eight example games.}
\label{ta:example-games}
\begin{tabular}{c c l l}
\toprule
Game & $K$ & SNEs & ESSs \\
\midrule
1 &
2 &
$\{(0,1),\; (1/3,2/3),\; (1/2,1/2)\}$ &
$\{(0,1),\; (1/2,1/2)\}$ \\
\midrule
2 &
2 &
All mixed strategies &
$\varnothing$ \\
\midrule
3 &
2 &
$\{(1,0),\; (0,1)\}$ &
$\{(1,0)\}$ \\
\midrule
4 &
3 &
$\{(1/3,\,1/3,\,1/3)\}$ &
$\varnothing$ \\
\midrule
5 &
3 &
$\{(1,0,0)\} \;\cup\; \{(0,x,y): x,y \ge 0,\ x+y=1\}$ &
$\{(1,0,0)\}$ \\
\midrule
6 &
3 &
$\{(1,0,0),\ (0,1,0),\ (0,0,1),\ (0.5, 0, 0.5)\}$ &
$\{(1,0,0),\ (0,0,1)\}$ \\
\midrule
7 &
3 &
\begin{tabular}[t]{@{}l@{}}
$\{(1,0,0),\ (0,1,0),\ (0,0,1),$ \\
$(3/7,\,4/7,\,0),\ (0,\,1/3,\,2/3),\ (0.2,\,0.4,\,0.4)\}$
\end{tabular}
&
$\{(1,0,0),\ (0,1,0)\}$ \\
\midrule
8 &
3 &
$\{(0.4311484,\,0.3760157,\,0.1928359)\}$ &
$\{(0.4311484,\,0.3760157,\,0.1928359)\}$ \\
\bottomrule
\end{tabular}
\end{table}

For our next experiments, we ran our algorithm on three-player symmetric normal-form games generated by drawing each independent payoff entry from $U(-1,1)$ and imposing $A[i,j,k]=A[i,k,j]$.
We generated and solved 100 games for each number of pure strategies $K$ from 3 to 8. We used the nonconvex quadratic solver from Gurobi version 13.0.3~\cite{Gurobi26:Gurobi}.
We used an Intel Core i7-1065G7 CPU with 16 GB of RAM, with Gurobi configured to use 8 threads. The SNE-search subproblems used \texttt{NumericFocus=3}, and the remaining numerical parameter values are described above and in Table~\ref{ta:parameters}. The pseudorandom seed was 12345, reset for each value of $K$.

The runtimes from the experiments are given in Table~\ref{ta:runtime}. For each $K$ we report the total runtime of Algorithm~\ref{al:main}, as well as the runtime until the first ESS is found. If no ESS is found, then we report the time to termination as the first-ESS time. For both the total and first runtimes we report the mean with 95\% confidence intervals as well as the median. For the biggest games $K = 8$ the mean total runtime is 15.14 seconds, and the mean runtime to find one ESS is 0.67 seconds. The mean total runtime is below one second for $K=3$--5. All 600 random-game runs completed without an unresolved ESS classification, incomplete support search, time-limit termination, or representative cap. The repeated support search found at least two separated SNE representatives on 23, 31, 50, 66, 80, and 97 supports for $K=3,4,5,6,7,8$, respectively.

\begin{table}[ht]
\centering
\caption{Runtime statistics over 100 random symmetric games for each number of pure strategies $K$.
Times reported in seconds. ``Total'' refers to the full numerical search.
``First'' is the time to find the first ESS (or terminate when none is found).}
\label{ta:runtime}
\begin{tabular}{c c c c c}
\toprule
$K$ &
\begin{tabular}{c}Mean Total \\ ($\pm$ 95\% CI)\end{tabular} &
\begin{tabular}{c}Median \\ Total\end{tabular} &
\begin{tabular}{c}Mean First \\ ($\pm$ 95\% CI)\end{tabular} &
\begin{tabular}{c}Median \\ First\end{tabular} \\
\midrule
3 & 0.0494 $\pm$ 0.0094 & 0.0404 & 0.0098 $\pm$ 0.0024 & 0.0029 \\
4 & 0.1644 $\pm$ 0.0369 & 0.1178 & 0.0175 $\pm$ 0.0080 & 0.0032 \\
5 & 0.5355 $\pm$ 0.0520 & 0.4843 & 0.0537 $\pm$ 0.0399 & 0.0045 \\
6 & 2.4540 $\pm$ 0.2212 & 2.2446 & 0.1005 $\pm$ 0.0795 & 0.0061 \\
7 & 6.4140 $\pm$ 0.6783 & 5.9255 & 0.7036 $\pm$ 0.7353 & 0.0059 \\
8 & 15.1372 $\pm$ 0.8201 & 14.5217 & 0.6699 $\pm$ 0.6324 & 0.0063 \\
\bottomrule
\end{tabular}
\end{table}

Table~\ref{ta:ess-hist} provides further insight by showing the number of ESSs 
found for each $K$ (aggregated over the 100 sampled games). While a small number
of games had no ESS, the majority of games had a moderate number of ESSs (1--3). 
Table~\ref{ta:ess-support-sizes} gives a breakdown of the number of ESSs found with each
support size. A large number of the ESSs had support 
size of 1 or 2, which explains why
the running times to find the first ESS were so low; however some games had ESSs with larger support
sizes.

\begin{table}[ht]
\centering
\caption{Histogram of ESS counts over 100 random symmetric games for each $K$.
Entry $(K,i)$ is the number of games with $i$ ESSs.}
\label{ta:ess-hist}
\begin{tabular}{c|cccccccc}
\toprule
$K$ &
0 & 1 & 2 & 3 & 4 & 5 & 6 & 7 \\
\midrule
3 & 3 & 37 & 49 & 11 & 0 & 0 & 0 & 0 \\
4 & 2 & 38 & 39 & 17 & 4 & 0 & 0 & 0 \\
5 & 5 & 23 & 36 & 23 & 11 & 2 & 0 & 0 \\
6 & 2 & 20 & 41 & 25 & 10 & 2 & 0 & 0 \\
7 & 4 & 20 & 33 & 25 & 11 & 4 & 2 & 1 \\
8 & 4 & 12 & 33 & 31 & 15 & 3 & 2 & 0 \\
\bottomrule
\end{tabular}
\end{table}

\begin{table}[ht]
\centering
\caption{ESS support-size distribution over 100 random symmetric games for each $K$.
Entry $(K,s)$ reports the total number of ESSs found with support size $s$ over all games with $K$ pure strategies.}
\label{ta:ess-support-sizes}
\begin{tabular}{c|ccccc}
\toprule
$K$ & $s{=}1$ & $s{=}2$ & $s{=}3$ & $s{=}4$ & $s{=}5$ \\
\midrule
3 & 101 & 63 & 4 & 0 & 0 \\
4 & 92 & 66 & 23 & 2 & 0 \\
5 & 91 & 103 & 22 & 2 & 0\\
6 & 102 & 76 & 44 & 5 & 0\\
7 & 101 & 104 & 29 & 9 & 1\\
8 & 109 & 94 & 50 & 4 & 1 \\
\bottomrule
\end{tabular}
\end{table}

Table~\ref{ta:ess-breakdown} provides a breakdown of the outcomes of the ESS test described in Algorithm~\ref{al:isess}. The failure columns indicate the stage at which candidate SNEs are eliminated. Thus,
\[
\text{Total ESSs}
=
\text{Total SNEs}
-
\text{Pure-Fail-L1}
-
\text{Pure-Fail-L2}
-
\text{Mixed-Fail-L1}
-
\text{Mixed-Fail-L2}.
\]
The results show that the pure-mutant screen eliminates the majority of non-strict SNEs. For example, when $K=8$, the algorithm encounters 1482 SNEs, of which 109 are immediately classified as ESSs by the strict NE shortcut. Among the remaining 1373 non-strict SNEs, the first-level pure-mutant test eliminates 1166 (and the second-level pure-mutant test eliminates one). Thus only 206 candidates require solving the mixed-mutant optimization problems, a reduction of 85.0\% relative to testing all non-strict SNEs by mixed-mutant optimization. In total, the strict shortcut and pure-mutant screen reduce the number of SNEs for which the mixed-mutant QCQP procedures must be applied from 1482 to 206, which is a reduction of 86.1\%.
The second-level pure-mutant test is rarely decisive in these random games, while the second-level mixed-mutant test eliminates a small but nonzero number of candidates. 

\setlength{\tabcolsep}{4pt}
\begin{table}[ht]
\centering
\caption{Breakdown of outcomes of the ESS test aggregated over 100 random symmetric $K$-strategy games.
``Total SNEs'' is the number of symmetric Nash equilibria encountered by the algorithm.
``Strict SNEs'' are equilibria immediately classified as ESSs via the strict Nash equilibrium shortcut.
``Pure-Fail-L1'' and ``Pure-Fail-L2'' are equilibria eliminated by the first- and second-level pure-mutant tests, respectively.
``Mixed-Fail-L1'' and ``Mixed-Fail-L2'' are equilibria eliminated by the first- and second-level mixed-mutant tests, respectively.
``Total ESSs'' is the number of equilibria that survive all ESS tests.}
\label{ta:ess-breakdown}
\begin{tabular}{|c|c|c|c|c|c|c|c|}
\hline
$K$ 
& Total SNEs
& Strict SNEs 
& Pure-Fail-L1
& Pure-Fail-L2 
& Mixed-Fail-L1 
& Mixed-Fail-L2 
& Total ESSs \\
\hline
3 &272 &101 &102 &0 &2 &0 &168 \\
4 &374 &92 &172 &0 &11 &8 &183 \\
5 &640 &91 &399 &0 &16 &7 &218 \\
6 &833 &102 &564 &0 &36 &6 &227 \\
7 &1102 &101 &801 &3 &46 &8 &244 \\
8 &1482 &109 &1166 &1 &47 &10 &258 \\
\hline
\end{tabular}
\end{table}

\FloatBarrier
\section{Conclusion}
\label{se:conclusion}
We introduced the first algorithm for computing evolutionarily stable strategies in symmetric multiplayer normal-form games. Our approach enumerates supports, computes symmetric Nash equilibria, and determines whether they are evolutionarily stable. The method combines two efficient preprocessing tests---the strict Nash equilibrium shortcut and a pure-mutant screen---with mixed-mutant quadratically-constrained quadratic programs for the remaining candidates. Experiments on eight example games and random games with up to eight strategies demonstrate that the algorithm is fast in practice. The breakdown of algorithmic components shows that the preprocessing tests reduce the number of SNEs requiring the mixed-mutant QCQP procedures by 86.1\% for $K = 8$, highlighting the practical impact of these refinements. Our results provide the first computational method for studying evolutionary stability in multiplayer interactions, with potential applications to evolutionary game theory, behavioral ecology, and tumor ecology. 

While the presentation of our algorithm and our experiments were focused on 3-player games, our algorithm is straightforwardly applicable to games with $n > 3$ players. As described in Section~\ref{se:algorithm}, the QCP for computing an SNE in Algorithm~\ref{al:sne-support} will involve expressions that are degree-($n-1$) polynomial in the variables instead of quadratic. We can convert these to quadratic expressions by adding auxiliary variables, e.g., $p_{12}=p_1p_2$, $p_{123}=p_{12}p_3$. In general, for support size $s$ and $n$ players, this requires $O(s^{n-1})$ variables, which is $O(K^{n-1})$ in the worst case where $K$ is the number of pure strategies. Thus, the number of variables is exponential in $n$, but polynomial in $K$ for fixed $n$. The $H_1$ mixed-mutant test remains quadratic for $n>3$, while the full ESS test requires additional higher-level mutant comparisons, which can likewise be reformulated as QCQPs using auxiliary product variables. The total number of supports enumerated over remains $2^K-1$ independently of $n$. By contrast, if we were iterating over all possible support combinations to find a Nash equilibrium in a game that is not necessarily symmetric, we would need to consider $(2^K-1)^n$ support profiles, which is exponential in both $K$ and $n$. 

While we presented several preprocessing procedures that are shown to significantly improve computational efficiency---the strict Nash equilibrium shortcut and pure-mutant screen---our algorithm's efficiency can likely be further optimized significantly. A support-enumeration approach for computing a Nash equilibrium in $n$-player normal-form games (that are not necessarily symmetric) performs a recursive procedure of iteratively removing conditionally strictly dominated strategies before solving a feasibility program~\cite{Porter08:Simple}. Perhaps this additional preprocessing procedure could be effective in reducing the number of supports considered and/or the size of the corresponding QCPs solved in our algorithm as well. With access to additional computational resources our algorithm could also benefit from parallel computation, most obviously by solving multiple QCPs or QCQPs simultaneously. We showed that for 3-player games with $K = 8$ strategies per player our algorithm is able to run quickly on a laptop without any of these additional enhancements. It is promising that our algorithm can be applied to small and moderate-sized games already in its current form. For larger games that the algorithm is unable to solve, it is still possible that our algorithm can solve them if augmented by efficiency enhancements such as those described above.

\bibliographystyle{plain}
\bibliography{C://FromBackup/Research/refs/dairefs}

\appendix
\section{Payoff Tensors for the Example Games}
\label{ap:examples}

All games are symmetric 3–player. For each action $i$, 
$A[i,:,:]$ gives the focal player's payoff when the co–players use $j,k$.

\paragraph{Game 1 (two–strategy game with pure and mixed ESSs)}
\[
A[0] =
\begin{pmatrix}
-3 & 0\\
0 & -3
\end{pmatrix},
\qquad
A[1] =
\begin{pmatrix}
1 & -3\\
-3 & -1
\end{pmatrix}.
\]

\paragraph{Game 2 (neutral two–strategy game)}
\[
A[0] =
\begin{pmatrix}
1 & 0\\
0 & 1
\end{pmatrix},
\qquad
A[1] =
\begin{pmatrix}
1 & 0\\
0 & 1
\end{pmatrix}.
\]

\paragraph{Game 3 (weak dominance of strategy 0)}
\[
A[0] =
\begin{pmatrix}
2 & 0\\
0 & 0
\end{pmatrix},
\qquad
A[1] =
\begin{pmatrix}
0 & 0\\
0 & 0
\end{pmatrix}.
\]

\paragraph{Game 4 (three–strategy rock–paper–scissors)}
\[
A[0] =
\begin{pmatrix}
0 & -1 & 1\\
-1 & -2 & 0\\
1 & 0 & 2
\end{pmatrix},
\quad
A[1] =
\begin{pmatrix}
2 & 1 & 0\\
1 & 0 & -1\\
0 & -1 & -2
\end{pmatrix},
\quad
A[2] =
\begin{pmatrix}
-2 & 0 & -1\\
0 & 2 & 1\\
-1 & 1 & 0
\end{pmatrix}.
\]

\paragraph{Game 5 (strict pure ESS at $(1,0,0)$)}
\[
A[0] =
\begin{pmatrix}
2 & 0 & 0\\
0 & 0 & 0\\
0 & 0 & 0
\end{pmatrix},
\qquad
A[1] =
\begin{pmatrix}
0 & 0 & 0\\
0 & 0 & 0\\
0 & 0 & 0
\end{pmatrix},
\quad
A[2] =
\begin{pmatrix}
0 & 0 & 0\\
0 & 0 & 0\\
0 & 0 & 0
\end{pmatrix}.
\]

\paragraph{Game 6 (Proliferator–Producer–Invasive tumor game)}
\[
A[0] =
\begin{pmatrix}
4 & 2 & 3\\
2 & 0 & 1\\
3 & 1 & 2
\end{pmatrix},
\quad
A[1] =
\begin{pmatrix}
0 & 0 & 0\\
0 & 0 & 0\\
0 & 0 & 0
\end{pmatrix},
\quad
A[2] =
\begin{pmatrix}
2 & 1 & 3\\
1 & 0 & 2\\
3 & 2 & 4
\end{pmatrix}.
\]

\paragraph{Game 7 (three–strategy game with six SNEs and two ESSs)}
\[
A[0] =
\begin{pmatrix}
6 & 4 & 5\\
4 & 2 & 3\\
5 & 3 & 4
\end{pmatrix},
\quad
A[1] =
\begin{pmatrix}
2 & 3.5 & 2.5\\
3.5 & 5 & 4\\
2.5 & 4 & 3
\end{pmatrix},
\quad
A[2] =
\begin{pmatrix}
4 & 3.5 & 4\\
3.5 & 3 & 3.5\\
4 & 3.5 & 4
\end{pmatrix}.
\]

\paragraph{Game 8 (random three–strategy game with mixed ESS)}
The six-decimal payoffs below are the values used in the implementation; the equilibrium in Table~\ref{ta:example-games} is rounded for display.
\[
A[0] =
\begin{pmatrix}
-1.317028 & -0.165184 & -0.549312\\
-0.165184 & 0.986660 & 0.602532\\
-0.549312 & 0.602532 & 0.218404
\end{pmatrix},
\]
\[
A[1] =
\begin{pmatrix}
0.986660 & -0.312191 & 0.559866\\
-0.312191 & -1.611041 & -0.738984\\
0.559866 & -0.738984 & 0.133073
\end{pmatrix},
\]
\[
A[2] =
\begin{pmatrix}
0.218404 & 0.175738 & -0.665863\\
0.175738 & 0.133073 & -0.708528\\
-0.665863 & -0.708528 & -1.550130
\end{pmatrix}.
\]

\end{document}